\documentclass[aps,prl,preprint,superscriptaddress,showpacs]{revtex4-1}
\usepackage{graphics}
\begin{document}

\preprint{}

\title{Surface Plasmon Polaritons Excited By Electromagnetic Waves Under General Boundary Conditions}


\author{Weiguo Yang}
\email{wyang@wcu.edu}
%\thanks{}
\affiliation{Department of Engineering \& Technology, Western Carolina University - Cullowhee, NC 28723, U. S. A. \\ Email: wyang@wcu.edu\\Tel: 828-227-2693  Fax: 828-227-3805}

\author{Michael A. Fiddy}
\affiliation{Center for Optoelectronics and Optical Communications, University of North Carolina at Charlotte - Charlotte, NC 28223, U. S. A.}


\date{\today}

\begin{abstract}
We investigate single interface surface plasmon polaritons (SPPs) excited by electromagnetic waves under general electromagnetic boundary conditions that allow for non-zero surface charge and surface current densities. Incorporating these, we derive general conditions for single interface SPPs and solve both the surface source distributions and the surface electromagnetic waves for such SPPs when excited by electromagnetic waves.   
%\\ \\
%Keywords: Applied classical electromagnetism, Reflection and refraction, Metamaterials, Plasmonics, boundary conditions
\end{abstract}

%{Applied classical electromagnetism}
%\pacs{42.25.Gy}%{Edge and boundary effects; reflection and refraction}
%\pacs{42.30.Va}%{Image forming and processing}

%\keywords{Applied classical electromagnetism, Metamaterials, Optical imaging, Reflection and refraction}

\maketitle

%\section{}
%\subsection{}
%\subsubsection{}

% If in two-column mode, this environment will change to single-column
% format so that long equations can be displayed. Use
%\begin{widetext}
%\end{widetext}

\section{Introduction}
General electromagnetic boundary conditions allow for non-zero surface charge and non-zero surface current densities \cite{JDJackson,rftext}. The only underlying assumption for such general boundary conditions is that there are no magnetic monopoles. However, in many applications, especially in optics, boundary conditions are enforced that require continuity of the tangential components of the electric field ${\bf E}$ and the normal component of the magnetic field ${\bf B}$ across the interface.  Continuity of the normal component of the of the electric displacement field ${\bf D}$ and tangential component of the magnetizing field, ${\bf H}$ is also usually assumed.  These boundary conditions are based on underlying assumptions that there are no surface charges or surface currents present. For a wide range of situations such surface sources can be safely assumed to be zero.  Active media and/or complex media interfaces often do require such surface sources to be present \cite{rftext, superlattice}. Equivalent surface sources are also introduced in the method of moments (MoM) to correctly describe the effects of external fields on field solutions in some region of interest\cite{39}.  The presence of a surface current leads to a discontinuity of the tangential ${\bf H}$ field and a surface charge density leads to the discontinuity in the normal component of ${\bf D}$. In this letter, we rigorously describe surface plasmon polaritons (SPPs) excited by electromagnetic waves at a single interface between two arbitrary media assuming the most general electromagnetic boundary conditions that allow for non-zero surface sources.   
 
\begin{figure}
\includegraphics{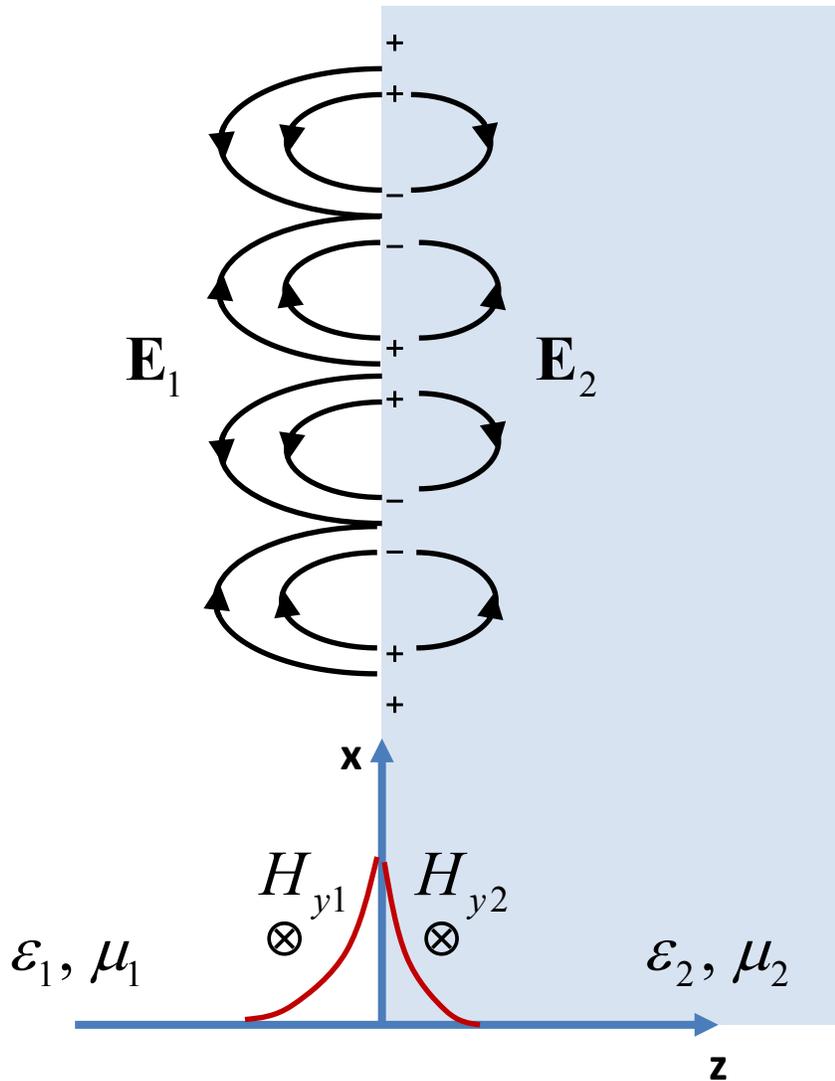}
\caption{Classic surface plasmon polariton: The surface charge density distribution and the associated electromagnetic fields including the exponential dependence of the tangential component of the magnetic fields on the distance away from the interface are illustrated.}
\label{fig1}
\end{figure}

The traditional description for SPPs comes from the derivation of surface modes of the electromagnetic field at an interface between a dielectric medium, for example air, and a metal; the usual electromagnetic boundary conditions are thus imposed that require continuity of the tangential components of both the electric field ${\bf E}$ and the magnetic field ${\bf H}$. Solutions representing the surface electromagnetic waves of the SPPs only exist and propagate along the boundary and decay in both media away from the interface as shown in Fig. \ref{fig1}.  
It has been long understood \cite{Agarwal_1973} that SPPs are self-sustained and can exist with zero input field. In this classical model, surface electromagnetic waves are a consequence of the discontinuous normal component of the electric field ${\bf E}$, but there is no surface current because the tangential components of the magnetic fields ${\bf H}$ on either side of the boundary remain continuous, i.e. there is no net surface current.  When a possible non-zero surface current is involved, the tangential component of the magnetic field ${\bf H}$ will become discontinuous across the boundary. Furthermore, when excited by non-zero input electromagnetic fields, SPPs will necessarily be accompanied by non-zero surface currents. When the input field is withdrawn, the resulting surface current also disappears and the SPPs take the traditional form and become self-sustained. The SPPs with non-zero surface current can also couple efficiently into the self-sustained SPPs in the region where there are no excitation fields.  In direct contradiction to the arguments that the SPPs cannot coexist with non-zero surface currents, we believe the self-consistent theory presented here offers a critical piece understanding the whole picture of SPP excitation and propagation.
 
\section{SPP excited by electromagnetic waves under general boundary conditions}

\begin{figure}
\includegraphics{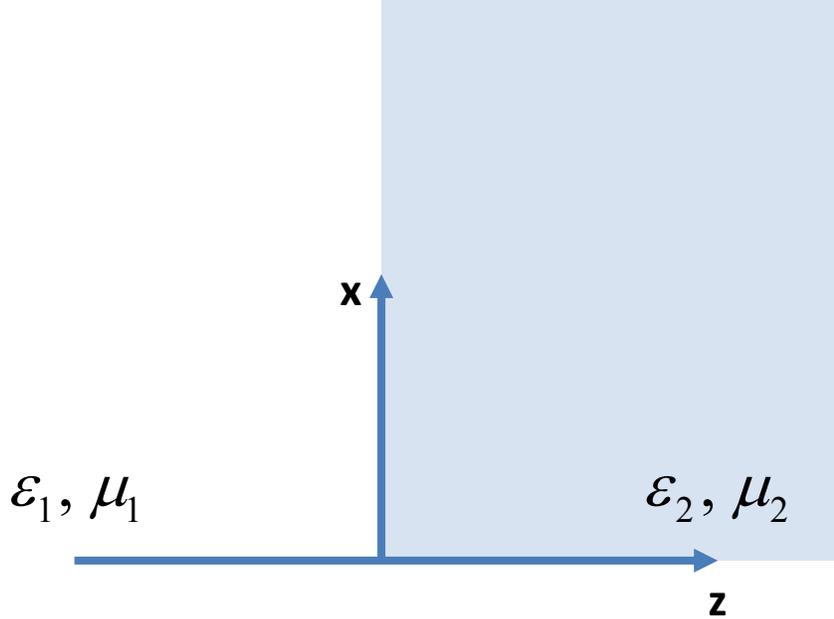}
\caption{Interface between two uniform media.}
\label{fig2}
\end{figure}

Consider the case of an interface between two uniform media with arbitrary relative permittivity $\epsilon_{1,2}$ and relative permeability $\mu_{1,2}$, where the input electromagnetic wave is incident from the left side ($z<0$) of the interface, as shown in Fig. \ref{fig2}. 

We first consider P-polarized waves for which the magnetic field of the input wave is given by,
\begin{equation}
{{\mathbf{H}}_{1P+}}=[0,1,0]{{H}_{0}}\exp (i{{k}_{1z}}z+i{{k}_{1x}}x-i\omega t),
\label{eq7}\end{equation}
where the wave vector component $k_{1z}=\pm\sqrt{\epsilon_1\mu_1k_0^2-k_x^2}$, $k_0=\omega/c$, and $c$ is the speed of light in vacuum.  The components of the wave vector in general can be complex. The choice of the positive or negative branch is usually determined by other considerations. For example, the non-zero real part of $k_z$ should take the same sign as the real part of index of the refraction of the medium. This follows because, for positive (real part) index media, the phase propagation should be in the same direction as the energy propagation, while for negative index media, the real part of $k_z$ should be negative so that the phase velocity is in the opposite direction to the direction of energy flow.

The magnetic field of the reflected wave is
\begin{equation}
{{\mathbf{H}}_{1P-}}= [0,1,0]{{H}_r}\exp (-i{{k}_{1z}}z+i{{k}_{1x}}x-i\omega t),
\label{eq8}\end{equation}
and the magnetic field of the transmitted waves in medium 2 is,
	\begin{equation}
{{\mathbf{H}}_{2P+}}= [0,1,0]{{H}_t}\exp (i{{k}_{2z}}z+i{{k}_{2x}}x-i\omega t),
\label{eq9}\end{equation}
where $H_r$ and $H_t$ are the amplitudes of the reflected and transmitted magnetic fields.  Depending on the index difference across the boundary, equation (3) can contain high-k, i.e. evanescent wave components. 

Assuming the bulk region of the medium is source free, Maxwell's equations give $-i\omega {\epsilon{\epsilon}_{0}}\mathbf{E}=\nabla \times \mathbf{H}$. Accordingly, assuming $\epsilon_1\neq0$ and $\epsilon_2\neq0$, one has for the electric field of the incident wave,
	\begin{equation}
{{\mathbf{E}}_{1P+}}=-\frac{H_0}{\omega\epsilon_1 \epsilon_0}[-k_{1z},0,k_{1x}] \exp (i{{k}_{1z}}z+i{{k}_{1x}}x-i\omega t),\label{eq10}
\end{equation}

for the electrical field of the reflected wave, 
	\begin{equation}
{{\mathbf{E}}_{1P-}}=-\frac{H_r}{\omega \epsilon_1\epsilon_0}[k_{1z},0,k_{1x}] \exp (-i{{k}_{1z}}z+i{{k}_{1x}}x-i\omega t),
\label{eq11}
\end{equation}

and for the electric field of the transmitted wave,
	\begin{equation}
{{\mathbf{E}}_{2P+}}=-\frac{H_t}{\omega\epsilon_2 \epsilon_0}[-k_{2z},0,k_{2x}] \exp (i{{k}_{2z}}z+i{{k}_{2x}}x-i\omega t).
\label{eq12}\end{equation}

Note that media with $\epsilon=0$ do not support P-polarized waves.

At the interface of two media, one has the general electromagnetic boundary conditions given below, which take into account the possibility of non-zero surface charge density $\Sigma$ and non-zero surface current density ${\bf K}$ \cite{JDJackson,rftext}, 

\begin{eqnarray}
{\bf n}\times ({\bf E}_2-{\bf E}_1)=0 \\
{\bf n}\cdot ({\bf D}_2-{\bf D}_1)=\Sigma \\
{\bf n}\times ({\bf H}_2-{\bf H}_1)={\bf K} \\
{\bf n}\cdot ({\bf B}_2-{\bf B}_1)=0  
\end{eqnarray}
where ${\bf n}$ is the unit normal vector pointing from medium 1 to medium 2.  The surface charge density and the surface current density  satisfy the continuity relation, following the continuity law of charge density and current density,

\begin{equation}
\nabla \cdot {\bf K} -i\omega\Sigma=0. 
\label{cont}
\end{equation}

The tangential component of the electric field ${\bf E}$ and the normal component of the magnetic field flux ${\bf B}$ remain continuous across the boundary because physically there is no magnetic monopole nor magnetic current.

Applying these physical boundary conditions at the interface where $z=0$, and assuming $k_{1z}\neq0$, one has the following for the reflected and transmitted fields $H_r$ and $H_t$ 

\begin{equation}
H_r=H_0-\frac{\epsilon_1k_{2z}}{\epsilon_2k_{1z}} H_t.
\label{eq_rt}
\end{equation}

For the surface charge densities, one has,
\begin{equation}
\Sigma=\frac{k_x}{\omega}\left[2H_0-\left(1+\frac{\epsilon_1k_{2z}}{\epsilon_2k_{1z}}\right) H_t\right] e^{ik_x x-i\omega t}.
\end{equation}

For the surface current densities,  $K_y=K_z=0$ and,

\begin{equation}
K_x= \left[2 H_0-\left(1+\frac{\epsilon_1k_{2z}}{\epsilon_2k_{1z}}\right)H_t\right]e^{ik_x x-i\omega t}.
\end{equation}

Obviously, the continuity condition Eq.(\ref{cont}) for $\Sigma$ and ${\bf K}$ is satisfied by these expressions.  Also, implicit here, is that $k_{1x}=k_{2x}=k_x$, which is the generalized Snell's law of refraction in the absence of a surface phase discontinuity.  For the cases where $1+\epsilon_1k_{2z}/\epsilon_2k_{1z} \neq0$, these boundary conditions support non-trivial field solutions for $H_t$ and $H_r$ when $\Sigma=0$ and ${\bf K}=0$, which results in the well-known Fresnel equations for the transmitted and reflected waves \cite{WolfBorn}. 

The cases where 

\begin{equation}
1+\frac{\epsilon_1k_{2z}}{\epsilon_2k_{1z}}=0
\label{condTM}
\end{equation}

can be easily recognized as the condition for SPP.  However, in such cases one definitely has non-zero surface charge and current densities whenever $H_0\neq 0$, i.e. we have 

\begin{equation}
\Sigma=\frac{2k_xH_0}{\omega} e^{ik_x x-i\omega t}
\end{equation}
and
\begin{equation}
K_x=2H_0 e^{ik_x x-i\omega t}
\end{equation}

The amplitude of SPPs for cases where $H_0\neq0$ can be calculated using the Greens function method. For the simple cases where the media are lossless, it is straightforward to show that $H_t=-H_0$ and $H_r=0$. 

A special case for SPPs exists when $H_0=0$. Obviously the surface charges and surface currents are zero under these circumstances, i.e. $\Sigma=0$ and ${\bf K}=0$. However, according to Eq.(\ref{eq_rt}) and Eq.(\ref{condTM}) one has,

\begin{equation}
H_r=H_t.
\end{equation}

This equality means that for the special SPP cases where the input $H_0=0$, there are no surface sources and SPPs of arbitrary amplitude $H_r=H_t$ can exist. Clearly, this is the case for our traditional understanding of SPPs. However, when there is non-zero input field then $H_0\neq0$, and non-zero surface charges and surface currents will be  called for under the condition for SPPs.  This scenario has not been extensively investigated so far and we argue that this clarifies how SPPs can coexist with non-zero surface currents.  

A similar analysis can be made for S-polarized waves for which the electric field is given by,
\begin{equation}
{{\mathbf{E}}_{1S+}}=[0,1,0]{{E}_{0}}\exp (i{{k}_{1z}}z+i{{k}_{1x}}x-i\omega t),
\label{eq1}
\end{equation}
the electric field of the reflected waves is
\begin{equation}
{{\mathbf{E}}_{1S-}}=[0,1,0]{{E}_r}\exp (-i{{k}_{1z}}z+i{{k}_{1x}}x-i\omega t),
\label{eq2}
\end{equation}
and the electric field of the transmitted waves is,
	\begin{equation}
{{\mathbf{E}}_{2S+}}=[0,1,0]{{E}_t}\exp (i{{k}_{2z}}z+i{{k}_{2x}}x-i\omega t)
\label{eq3}
.\end{equation}

Maxwell's equations relate the electric field to the magnetic field in the bulk of media where there are no free sources by $i\omega {\mu{\mu }_{0}}\mathbf{H}=\nabla \times \mathbf{E}$. Assuming $\mu_1\neq0$ and $\mu_2\neq0$ and applying the general boundary conditions at the interface where $z=0$, one obtains $\Sigma=0$, $K_x=K_z=0$, and

\begin{equation}
E_0+E_r=E_t
\label{eq_rtE} 
\end{equation}

and

\begin{equation}
K_y=\frac{1}{\omega \mu_0} \left[\frac{2k_{1z}}{\mu_1 }E_0-\left(\frac{k_{1z}}{\mu_1 }+\frac{k_{2z}}{\mu_2 }\right)E_t\right]e^{ik_x x-i\omega t}
\end{equation}

where the generalized Snell's law of refraction  $k_{1x}=k_{2x}=k_x$ has been applied. Note that media with $\mu=0$ do not support S-polarized waves.

In a similar manner, for the case where $k_{1z}/\mu_1 +k_{2z}/\mu_2 \neq0$, the boundary conditions support non-trivial field solutions of transmitted and reflected fields $E_t$ and $E_r$ when there are zero surface charges and currents. This corresponds to the well-known Fresnel equations for the transmitted and reflected waves. However, the case when 

\begin{equation}
\frac{k_{1z}}{\mu_1 }+\frac{k_{2z}}{\mu_2 }=0
\label{condTE}
\end{equation}

can be interpreted as the condition for S-polarized SPPs. Such condition is typically not satisfied for materials like dielectrics and other non-magnetic materials, and S-polarized SPPs do not exist for such materials. However, with the advancement in metamaterials, more exotic values of effective permittivity and permeability become possible. It is therefore possible for metamaterials to satisfy this condition for S-polarized SPPs, i.e. satisfy Eq.(\ref{condTE}) and hence support S-polarized SPPs.   

For this S-polarized SPP case,

\begin{equation}
K_y=\frac{2k_{1z}E_0}{\omega \mu_0\mu_1} e^{ik_x x-i\omega t}
\label{eq-ky}
\end{equation}
which is non-zero if $k_{1z}\neq 0$ and $E_0\neq 0$.

Similarly, the amplitude of S-polarized SPPs for cases where $E_0\neq0$ can be calculated using the Green's function method. For the simple cases where the media are lossless, it is straightforward to show that $E_t=E_0$ and $E_r=0$.  Also similar to the P-polarized example, the special case when there is no input field, i.e., $E_0=0$, can still be recognized.  From Eq.(\ref{eq_rtE}) and Eq.(\ref{eq-ky}) it immediately follows that when $E_0=0$, then there are no surface charges or currents and SPPs of arbitrary amplitude where $E_r=E_t$ can exist.

\section{Conclusions}
\label{exp}

We investigated in this paper SPPs excited by electromagnetic waves at a single interface assuming the most general electromagnetic boundary conditions that allow for non-zero surface charge and surface current densities.  We found that SPPs exist with or without surface charge and current distributions depending on whether there is a non-zero input field or not.  The SPPs can be self-sustaining without a stimulating non-zero input field, which was well known.  With a stimulating non-zero input field, however, SPPs in general coexist with and are inextricably linked to non-zero surface currents.  We derived the general conditions for single interface SPPs for both P- and S-polarized incident waves.  We also solved both the surface source distributions as well as the surface electromagnetic waves for both types of SPPs.

\newpage
\section*{Figure caption list}
Fig. \ref{fig1}: Classic surface plasmon polariton: The surface charge density distribution and the associated electromagnetic fields including the exponential dependence of the tangential component of the magnetic fields on the distance away from the interface are illustrated.

Fig. \ref{fig2}: Interface between two uniform media.
\end{document}